\documentclass[sn-basic]{sn-jnl}

\usepackage{graphicx}%
\usepackage{multirow}%
\usepackage{amsmath,amssymb,amsfonts}%
\usepackage{amsthm}%
\usepackage{mathrsfs}%
\usepackage[title]{appendix}%
\usepackage{xcolor}%
\usepackage{textcomp}%
\usepackage{manyfoot}%
\usepackage{booktabs}%
\usepackage{algorithm}%
\usepackage{algorithmicx}%
\usepackage{algpseudocode}%
\usepackage{listings}%
\usepackage{natbib}

\begin{document}

\title[Scaffold or Crutch? Examining College Students’ Use and Views of Generative AI Tools for STEM Education]{Scaffold or Crutch? Examining College Students’ Use and Views of Generative AI Tools for STEM Education}


\author*[1]{\fnm{Karen D.} \sur{Wang}}\email{kdwang@stanford.edu}
\author[1]{\fnm{Zhangyang} \sur{Wu}}
\author[1]{\fnm{L’Nard} \sur{Tufts II}}
\author[1]{\fnm{Carl} \sur{Wieman}}
\author[1]{\fnm{Shima} \sur{Salehi}}
\author[1]{\fnm{Nick} \sur{Haber}}
\affil[1]{\orgname{Stanford University}, \orgaddress{\city{Stanford}, \state{California} \postcode{94305}, \country{USA}}}

\newpage
\abstract{\textbf{Background:} Developing problem-solving competency is central to Science, Technology, Engineering, and Mathematics (STEM) education, yet translating this priority into effective approaches to problem-solving instruction and assessment remain a significant challenge. The recent proliferation of generative artificial intelligence (genAI) tools like ChatGPT in higher education introduces new considerations: how to define problem-solving competency in a genAI era, and how these tools can help or hinder students’ development of STEM problem-solving competency. Our research takes steps in examining these considerations by studying how and why college students are currently using genAI tools in their STEM coursework, with a specific focus on how they employ these tools to support their problem-solving. 

\par \textbf{Results:} We conducted an online survey of 40 STEM college students from diverse institutions across the US. In addition, we surveyed 28 STEM faculty to understand instructor views on effective and ineffective genAI tool use in STEM courses and their guidance for students. Our findings reveal high adoption rates and diverse applications of genAI tools among STEM students. The most common use cases of genAI tools in STEM coursework include finding explanations, exploring related topics, summarizing readings, and helping with problem-set questions. The primary motivation for using genAI tools in STEM coursework was to save time. Moreover, we found that over half of the student participants reported simply inputting a problem for AI to generate solutions, potentially bypassing their own problem-solving processes. 

\par \textbf{Conclusions:} These findings indicate that despite high adoption rates, students’ current approaches to utilizing genAI tools often fall short in enhancing their own STEM problem-solving competencies. The study also explored students’ and STEM instructors’ perceptions of the benefits and risks associated with using genAI tools in STEM education. Our findings provide insights into how to guide students on appropriate genAI use in STEM courses and how to design genAI-based tools to foster students’ problem-solving competency.}

\keywords{Generative AI, Educational Technology, STEM Education, Higher Education, Physics Education, Problem Solving}



\maketitle

\section{Introduction}\label{sec1}

The public debut of ChatGPT in 2022 has catalyzed adoption of Large Language Models (LLMs) and other genAI technologies in various fields. Within education, there is a growing recognition that AI will significantly impact not only how we teach and learn \citep{NEA2024AI}, but also what we should teach students to prepare them for a future where AI is increasingly integrated in daily tasks \citep{shiohira2021understanding}. The potential for genAI to disrupt established educational practices is especially pronounced in college STEM education for several reasons. First, developing problem-solving competency is one of the major goals in STEM education \citep{honey2020stem, abet2022criteria}, yet teaching, practicing and assessing this competency effectively and authentically have been challenging given the common instructional practices in STEM courses \citep{felder2012engineering}. These disciplines have traditionally relied on well-defined problems in homework assignments and exams to develop and assess student conceptual understanding and problem-solving competency \citep{jonassen2015all, montgomery2024characterizing}. However, LLMs have demonstrated high proficiency in solving these types of problems \citep{achiam2023gpt, west2023advances}, potentially altering students’ study habits, problem-solving approaches, and engagements with course materials. This capability of LLMs may thus render traditional instructional practices even less effective for developing and measuring STEM problem-solving competency. Second, as AI tools increasingly handle routine, well-defined tasks in technology-driven fields \citep{levy2013dancing}, STEM education must evolve to develop students’ competencies in effectively leveraging AI as a tool to solve authentic, complex problems that require human judgment and decision making. This shift is critical for preparing college graduates for the future of an AI-augmented workforce. 

\par Building on our extensive empirical research in improving authentic problem-solving education in college STEM courses \citep{salehi2018improving, burkholder2020template, price2022accurate}, we now examine the emerging influence and potential of genAI tools on students’ learning experiences in STEM coursework, with a focus on how these tools are affecting students’ problem-solving approaches. We seek to contribute to the ongoing discourse surrounding genAI’s impact on education by exploring the extent and ways in which college students are using genAI tools, what training they need to leverage these tools effectively for their learning and to prepare them for a future where AI is an integral part of it. Specifically, we propose the following research questions:

\textbf{RQ1:} How are college students currently using genAI tools generally and in their STEM coursework?

\textbf{RQ2:} What are students’ prompting behaviors when using genAI tools to support their STEM problem-solving?

\textbf{RQ3:} How do students rate the helpfulness of genAI tools in supporting various aspects of STEM problem-solving, and how do these ratings compare with faculty recommendations?

\textbf{RQ4:} What are the main benefits and risks perceived by students and faculty in using genAI tools in STEM education?

\section{Background}\label{sec2}

\subsection{Technology’s Role in Education}
Over the past few decades, the education field has witnessed waves of technological innovations, from Skinner’s teaching machine in the 1950s to the more recent emergence of Massive Online Open Courses (MOOCs) and adaptive computer-assisted instructional programs \citep{reich2020failure}. Research on educational technologies has shown mixed results: while some studies found positive effects of technology programs such as one-to-one laptops on academic achievement \citep{zheng2016learning}, others documented limited effectiveness or even negative consequences \citep{mueller2014pen, carter2017impact}. These results highlight the complexity of technology integration in education \citep{cuban2001oversold}. As Roy Pea on his examination of technology’s impacts on human cognition noted, technologies are not merely tools we use, but have the potential to fundamentally alter our behavior and cognitive processes, hence reorganize how we think, teach, and learn \citep{pea1985beyond}. With the emergence of genAI, we are once again facing the question of how this new technology may impact human learning, and in return educational practices, in ways both promising and troubling. 

\par Generative AI technologies, such as LLMs, are advanced AI systems pre-trained on simple tasks like predicting the next word in a text sequence \citep{brown2020languagemodelsfewshotlearners}. Through fine-tuning and prompting methods \citep{liu2022fewshotparameterefficientfinetuningbetter, wei2022finetunedlanguagemodelszeroshot}, these models have evolved beyond next-token prediction to unlocking a wide range of sophisticated capabilities, including engaging in real-time conversations as chatbots \citep{thoppilan2022lamda}, summarizing complex texts \citep{zhang2020pegasus}, and generating writings in diverse genres \citep{brown2020languagemodelsfewshotlearners}. Generative AI technologies such as LLMs have the potential to transform teaching and learning practices through automating the creation of instructional materials, enhancing assessment methods, and providing readily available personalized support to students \citep{kasneci2023chatgpt}. At the same time, the capability of LLMs to instantly generate contextually relevant and coherent content across various subjects poses a risk of students using them to complete coursework without genuine comprehension as a result of bypassing critical learning processes. 

\par A number of studies have surveyed college instructors and students to understand their perspectives and experiences with genAI \citep{amani2023generativeaiperceptionssurvey, goldberg2024generative, baek2024chatgpt}. These studies often highlight the duality of opportunities and risks associated with using genAI tools in education. For example, a survey for university instructors and students in the US and Australia revealed moderate usage of genAI tools for coursework and professional purposes \citep{10.1145/3573051.3596191}. At the same time, the study highlighted a shared concern among instructors and students that genAI would significantly impact certain types of assessments, including essays, computer codes, and short-answer questions. Another survey found that undergraduate and postgraduate students generally held positive attitudes towards genAI in teaching and learning and recognized several benefits associated with using genAI tools, such as personalized support and research assistance \citep{chan2023students}. However, students also expressed concerns about these tools’ accuracy, privacy, and impact on personal development and career prospects. These studies have largely examined the broad applications of genAI in educational contexts, yet there remains limited understanding of how discipline-specific characteristics interact with genAI capabilities to impact student learning in particular fields. 

\subsection{GenAI’s Role in Discipline-specific Education}
Recent studies have begun to examine how genAI tools can be effectively integrated to support the unique learning activities and pedagogical goals in different disciplines. In language instruction, a study involving college instructors found significant interest in adopting genAI to provide personalized feedback for students at different language proficiency levels, especially in grammar and pronunciation \citep{kohnke2023exploring}. At the same time, instructors recognized the urgency to address potential academic dishonesty issues arising from AI-generated writing. In medical education, recent research has identified promising applications of genAI such as self-directed learning, patient simulation, and writing assistance, while highlighting challenges such as maintaining academic integrity and ensuring data accuracy \citep{preiksaitis2023opportunities}. In computer science education, a research project combined a systematic literature review with a survey of instructors and students from 20 countries to examine the use of genAI tools such as Copilot \citep{10.1145/3623762.3633499}. The findings suggest that genAI tools could potentially enhance instructor productivity by aiding in the creation of instructional materials and automating grading and feedback in programming courses. On the other hand, the study also highlighted significant concerns, including students becoming overly dependent on these tools for coding tasks and potential academic integrity issues due to student submission of AI-generated solutions. These concerns are not unique to programming courses. Across disciplines, the pitfalls associated with genAI tools often share common elements of compromising academic integrity and/or bypassing critical learning processes, particularly related to more complex cognitive constructs such as problem-solving. 

\subsection{GenAI’s Role in STEM Education} 
The applications and implications of genAI tools in science education have received comparatively less attention than other fields. A study analyzing genAI-related policies in US higher education found a significant imbalance in the guidance provided for different disciplines \citep{mcdonald2024generative}. While the majority of R1 institutions (116 out of 131) had genAI-related policies as of November 2023, the more prevalent fields for their policies were in writing and humanities; only half of the institutions (n = 58) addressed the use of genAI in STEM courses. Within STEM courses, most policy mentions were in the context of computer science (n = 56), with disciplines like mathematics, natural science, and engineering receiving minimal attention. Furthermore, even when uses in STEM courses were addressed, the guidance was often superficial, primarily focusing on genAI’s roles in supporting coding and writing tasks rather than providing discipline-specific guidance. 

\par While institutional guidance on genAI usage in STEM education remains limited, researchers in Discipline-Based Education Research (DBER) have begun to investigate the application of genAI tools in STEM teaching and learning. One strand of the research focused on evaluating genAI’s capabilities in solving STEM problems and passing course exams \citep{kortemeyer2023could, hallal2023exploring, wang2024examining}. Another strand focused on exploring the applications of genAI tools in assisting instructors with creating instructional and assessment materials, grading student work, and providing feedback to students \citep{wan2024exploring, feldman2024perspectives}. Beyond these two main research strands, there is a growing body of work examining how students are using genAI tools in STEM courses, yielding both intriguing and sometimes concerning results. For instance, \cite{tassoti2024assessment} examined the prompting strategies of undergraduate students in a chemistry course and found that students predominantly relied on copying-pasting problems as their prompting strategy when interacting with genAI-based chatbots. In an introductory physics course, \cite{ding2023students} found that nearly half of the students trusted ChatGPT’s answers to physics problems regardless of their accuracy, with those expressing trust being more likely to report intentions of future ChatGPT use. Another study by \cite{bastani2024generative} found a complex relationship between access to generative AI tools and student performance in mathematics. While students with access to GPT-4 showed significant improvement on math practice problems, when the access was subsequently removed, they performed 17\% worse on exams compared to those who never had access. The authors proposed that this underperformance might be attributed to students becoming overly reliant on the genAI tool during practice sessions, using it to directly generate solutions without fully engaging with or understanding the material. These findings provide preliminary evidence that genAI tools, if not used effectively, can short-circuit students’ learning process. This underscores the need for more nuanced investigations into: 1) how students use genAI tools across various tasks in their STEM coursework and 2) what constitutes effective usage of these tools in STEM coursework. 

\par Our study aims to address this gap in research by understanding STEM students’ current usage patterns and perceived benefits and risks associated with using genAI tools in STEM coursework as well as STEM instructors’ guidance for students. We are particularly interested in exploring how genAI tools can support or hinder students’ learning of problem-solving, a broadly agreed-upon, integral goal for STEM education.  Our previous work has distilled a set of effective practices adopted and key decisions considered by experts in science and engineering domains during problem-solving \citep{salehi2018improving, price2021detailed}. Additionally, we have developed instructional materials to measure and teach authentic problem-solving skills to college students in STEM courses \citep{burkholder2020template, schwartz2024introducing}. Drawing from this research foundation, we now explore how genAI tools are, and can be, used for different aspects of STEM problem-solving. These aspects, including 1) identifying the relevant domain knowledge underlying a problem, 2) collecting the data need for solving a problem, 3) executing the problem-solving plan, and 4) verifying the correctness of a solution, have been empirically identified and validated in our previously developed problem-solving frameworks \citep{salehi2018improving}. Such exploration is crucial given the importance of helping students develop competency at solving novel authentic problems to prepare them for a future where AI increasingly automates routine tasks. We also compare students’ approach in using genAI for their STEM problem-solving with faculty recommendations on whether and how to use genAI tools to enhance learning STEM problem-solving. This comparison will provide valuable insights into the potential gaps between student practices and instructor guidance. Ultimately, we hope to identify promising directions for developing genAI-based tools to enhance students’ problem-solving competencies and create practical guidelines for students to effectively use these tools to enhance their learning outcomes in STEM disciplines.

\section{Methods}\label{sec3}
\subsection{Survey Design}
To investigate students’ usage and perceptions of genAI technologies for their STEM coursework, we developed an online survey through an iterative process involving reviews by STEM instructors and experts in STEM education and AI in education, as well as think-aloud pilot testing with college students. The final student survey includes five main components as listed in Table~\ref{tab1}. A parallel survey was designed to gather complementary perspectives from STEM instructors regarding genAI tools’ use in STEM courses and their recommendations for students.

\begin{table}[h]
\caption{Key constructs and components of the student survey}\label{tab1}%
\begin{tabular}{@{}llll@{}}
\toprule
Construct & Details & No. of Questions\\
\midrule
Usage    & General use of genAI tools and specific use in & 20   \\
&  STEM coursework, including names of tools, \\
& time of adoption, and primary reason for use \\ \midrule
Prompt Engineering Skills    & A prompt writing task that presents students    & 1
 \\
 & with a physics problem and asks them to \\
 & describe what prompt(s)/question(s) they \\
 & would give to a genAI-based chatbot to help\\
 & them with the problem \\ \midrule
Perceived Helpfulness    & Perceived helpfulness of genAI tools for various & 4\\
in Problem-solving & aspects of problem-solving in science and \\
& engineering disciplines  \\ \midrule
Perceived Benefits   & Perceived benefits and risks associated with  & 3 \\
and Risks & using genAI tools in STEM education \\ \midrule
Demographics and  & Participants’ academic year, field of study, & 8 \\ Academic Background & physics education background, and  \\ & demographic details \\

\botrule
\end{tabular}
\end{table}

\subsection{Data Collection}
Data collection for this study was conducted in two phases. In November 2023, we recruited faculty participants from the American Physical Society (APS) listserv and Slack channels. In May 2024, we launched the survey for college students studying STEM subjects using an online crowdsourcing platform, Prolific \citep{wang2024can}. The following filters were used to prescreen Prolific participants: 1) current undergraduate students; 2) located in the US; 3) studying one of the following STEM subjects: biochemistry, biomedical sciences, chemistry, engineering, materials science, mathematics, and physics. We did not recruit computer science faculty or students for this study because the use of genAI tools (e.g., Copilot) in computer science education has been more extensively studied compared to other STEM disciplines. 

\subsection{Data Analysis}
To address our research questions, we applied a combination of quantitative analyses to aggregate the data on participants’ usage and qualitative analyses to examine individual responses on their prompts used for problem-solving and perceived benefits and risks associated with using such technology in STEM education. For the qualitative analyses, we first conducted an open coding to examine all participants’ responses to specific questions for emerging themes. Following this, we established a set of codes accompanied by succinct, clear definitions and applied the coding framework to analyze individual responses. For all open-ended responses, two researchers independently coded at least 20\% of the data, with different pairs of researchers involved across various questions in the student and faculty survey. We calculated the agreement between two coders to assess inter-rater reliability for each of the questions analyzed. The mean agreement was 83.4\%, indicating generally high consistency in coding. Cases of disagreement were resolved through discussion, leading to refinement of code definitions when necessary. All quantitative data analyses were conducted using RStudio. 

\section{Results}\label{sec4}
Our analysis of survey data from 40 STEM college students and 28 physics instructors yielded insights into the current use, problem-solving approaches, perceived helpfulness, and potential impacts of genAI tools in STEM teaching and learning. We begin the results section with an overview of participant demographics to provide context. We then explore how and why students are using genAI tools generally and in their STEM coursework. This is followed by an examination of students’ prompting behaviors when applying these tools to support their STEM problem-solving. Next, we present students’ perceptions of genAI tools’ helpfulness across various aspects of STEM problem-solving and compare these views with instructor recommendations. Finally, we present an analysis of the primary benefits and risks associated with genAI use in STEM education, as perceived by both students and instructors. 

\subsection{Participant Demographics}
Our student participants came from diverse backgrounds, with 32.5\% identifying as female and 55\% as students of color. The majority of participants (n = 38) were enrolled in four-year universities, with one participant each from a community college and a vocational school. Participants also spanned all undergraduate years, with 3 freshmen, 7 sophomores, 15 juniors, and 15 seniors. Our instructor sample included 28 physics faculty with diverse levels of teaching experiences: 7 have taught for 4 years or less, 6 for 5-14 years, and 15 for 15 years or more. Faculty participants were predominantly from four-year universities and liberal arts colleges (n = 25), with the others from community colleges and a high school. Geographically, all student participants were based in the US. Among faculty participants, 24 were from North America, while 2 each were from Asia and Europe. 

\subsection{General Use of GenAI Tools}
\begin{figure}[h]
\centering
\caption{Student responses to “Which of the following generative AI tools have you used? Select all that apply.”
}\label{fig1}
\includegraphics[width=0.9\textwidth]{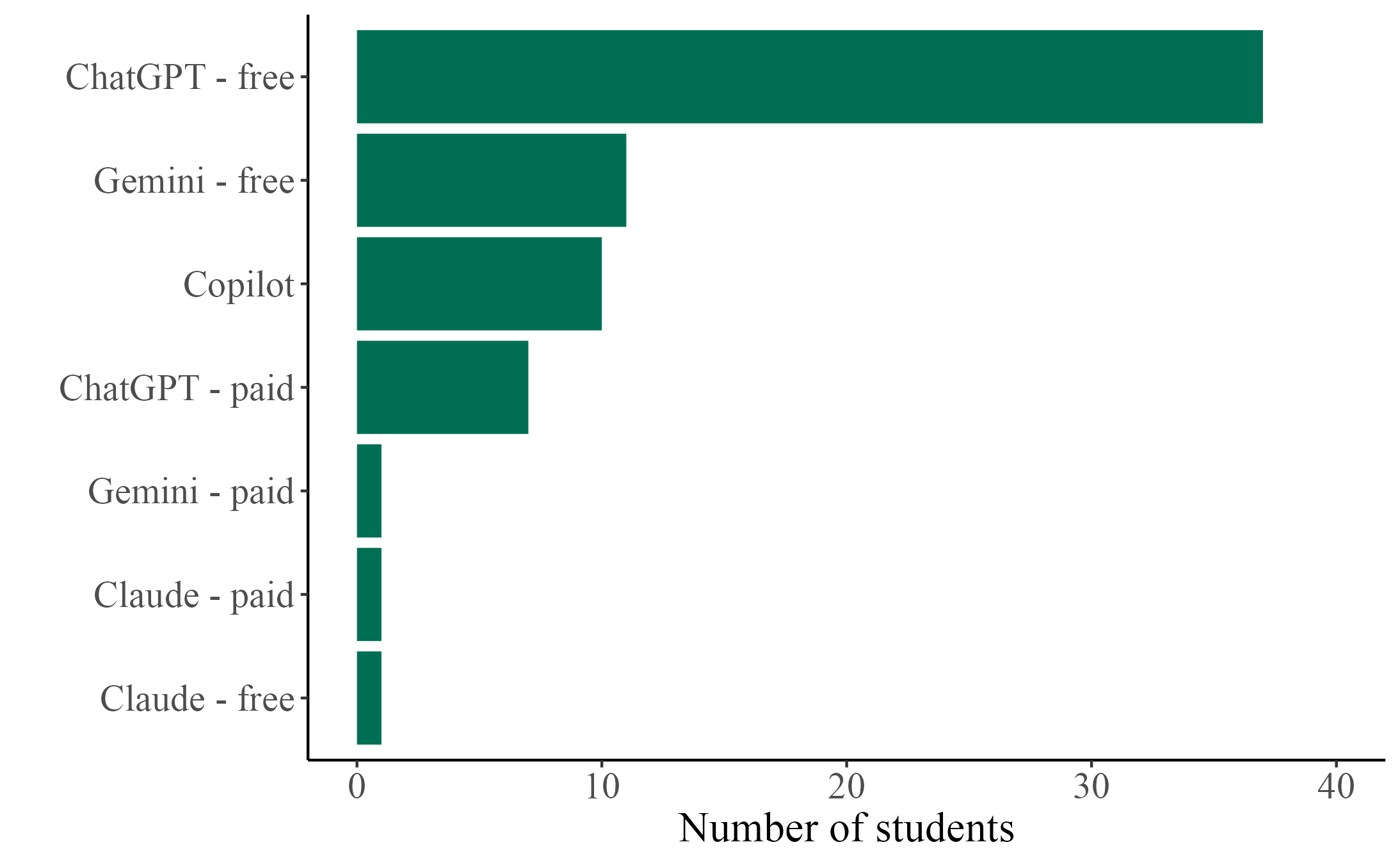}
\end{figure}

College students in our sample reported high adoption rates for genAI tools. Only 3 students (7.5\%) reported that they had not used any genAI tools as of May 2024. Regarding the adoption timeline, 4 students (10\%) began using such tools in 2022 or earlier, 29 (72.5\%) in 2023, and 4 (10\%) in 2024. In terms of specific tools, the free version of ChatGPT was the most widely used tool, with 37 students reporting its use. This was followed by the free versions of Gemini and Copilot, used by around a quarter of students each (n = 11 and 10, respectively). Paid versions of these tools saw significantly lower adoption rates. Fig.~\ref{fig1} presents the count of students using each of these tools. These results show that students were more likely to access the free versions of frontier LLMs, indicating that financial cost was one factor impacting college students’ use of genAI tools. 

\subsection{GenAI Use in STEM Coursework}
\begin{figure}[h]
\centering
\caption{Student responses to “How frequently do you use generative AI tools for the following tasks in your STEM (science, technology, engineering, and math) courses?”
}\label{fig2}
\includegraphics[width=\textwidth]{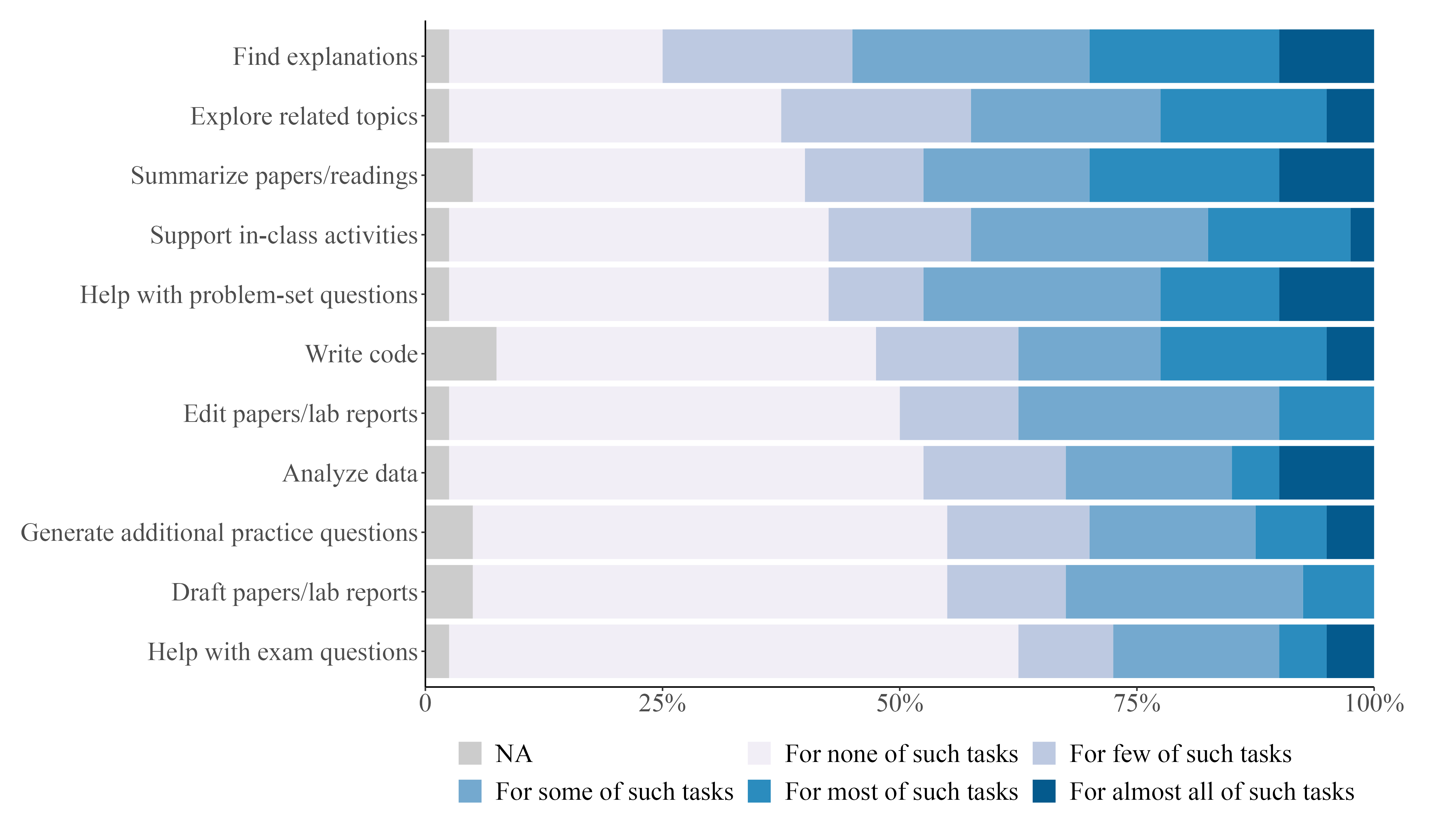}
\end{figure}

In the context of STEM coursework, our survey found that while 6 students reported never using genAI tools in their STEM courses, the remaining 34 students (85\%) utilized these tools for a variety of tasks as summarized in Fig.~\ref{fig2}. Using genAI tools to find explanations was the most common use, reported by 30 students (75\%). This was followed by exploring related topics (n = 25, 62.5\%), summarizing papers/readings (n = 24, 60\%), supporting in-class activities (n = 23, 57.5\%), and helping with problem-set questions (n = 23, 57.5\%). We did not find any significant association between students’ year in college and their usage patterns. When asked about their main reason for using genAI tools in STEM courses, the majority of students (n = 20, 50\%) indicated that it saved them time. The next most common reasons were the ability to ask questions without feeling judged and the accessibility of these tools, each cited by 5 students (12.5\%). Table~\ref{tab2} presents the detailed answer options for why using genAI tools, their corresponding constructs, and participants’ response frequencies. One student selected the “Other (please specify)” option and reported that they only used genAI tools when explicitly instructed to do so in their STEM courses and chose not to use them otherwise.  

\begin{table}[h]
\caption{Student responses to “What is your main reason for using generative AI tools in STEM courses? (Select one)” }\label{tab2}%
\begin{tabular}{@{}llll@{}}
\toprule
Answer Options & Underlying Construct & No. of Students\\
\midrule
It saves me time. & Efficiency & 20 (50\%) \\ \midrule
I can ask any question without feeling being judged. & Psychological safety & 5 (12.5\%) \\ \midrule
It is readily available. & Accessibility & 5 (12.5\%) \\ \midrule
It generates high-quality content. & High-quality & 3 (7.5\%) \\ \midrule 
Other (please specify:) & - & 1 (2.5\%) \\ \midrule 
I don’t use genAI tools in any of my STEM courses. & - & 6 (15\%) \\
\botrule
\end{tabular}
\end{table}

\begin{figure}[h]
\centering
\caption{Instructor responses to “How frequently do you use generative AI tools like ChatGPT for the following teaching tasks?”
}\label{fig3}
\includegraphics[width=\textwidth]{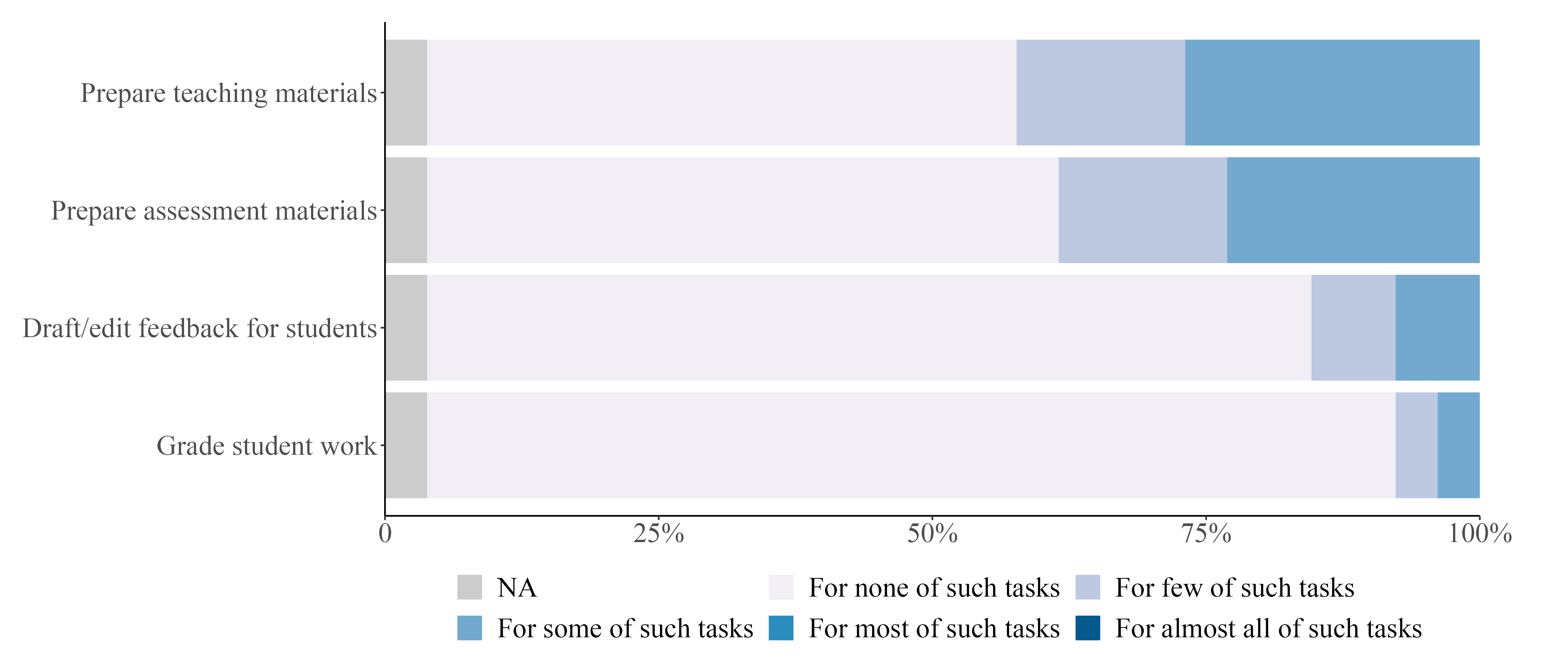}
\end{figure}

In contrast, the faculty members in our study exhibited a more cautious approach towards adopting genAI for teaching purposes. Fig.~\ref{fig3} presents faculty adoption rates of genAI tools for a range of teaching tasks as of November 2023. The majority of faculty members reported not using genAI for any of the teaching tasks listed, though a moderate number have begun to explore using such tools for preparing teaching materials (n = 11, 39\%) and assessment materials (n = 10, 36\%). On the other hand, faculty members were less likely to report adopting such tools for directly student-facing tasks, such as grading student work or providing feedback to students.  This was likely due to concerns about AI misinformation, which was a concern mentioned by a majority of the faculty. This pattern suggests that students are less likely to encounter genAI-based learning experiences initiated by their STEM faculty. However, this situation may evolve as genAI technology becomes integrated into familiar classroom tools and as the technology itself becomes more powerful, accessible, and trustworthy. We also found no significant correlation between faculty’s years of teaching experience and the adoption of genAI tools for teaching purposes.   

\subsection{Prompting Behaviors for STEM Problem-Solving}
To explore how students are using genAI tools to help with problem-solving in their STEM coursework, we gave students a complex problem from an introductory physics course and asked what prompts or questions they would give to a genAI tool like ChatGPT to help them with the problem. Qualitative analyses of their open-ended responses revealed that students demonstrated varying levels of prompt behaviors and epistemic beliefs for how genAI tools should be used to support their problem-solving. Table~\ref{tab3} presents the various approaches students adopted for prompting the genAI-based chatbot and the number and percentage of students whose responses fell into each category. 

\begin{table}[h]
\caption{Coded prompting strategies for using genAI in problem-solving based on student responses to “What prompt(s)/question(s) would you give to a generative AI tool like ChatGPT to help you with the problem?”}\label{tab3}%
\begin{tabular}{@{}llll@{}}
\toprule
Type of prompt & Number of students & Percentage\\
\midrule
Copy/paste or paraphrase the problem and ask AI to solve & 14 & 38\% \\ \midrule
Copy/paste or paraphrase the problem and provide & 6 & 16\% \\ 
additional instructions for AI to solve & & \\ \midrule
Ask specific questions that do not directly lead to a  & 17 & 46\% \\ 
solution but could help students figure it out & & \\
\botrule
\end{tabular}
\end{table}

The slight majority of students (54\%) in our sample preferred to use genAI to directly solve problems rather than as a tool to support their own problem-solving. This is evidenced by 38\% of students who would simply copy/paste or paraphrase the problem and ask AI to solve it, and an additional 16\% who would provide the problem along with some additional instructions for AI to solve. Examples of the additional instructions that students gave to genAI include “please support your answer,” and “take into account gravity, friction, and the average weight of the human body to calculate the time it takes to go from the bottom to the top floor.” In contrast, 46\% of the students asked specific questions that do not directly lead to a solution, suggesting a preference for guided support. Examples of the specific questions that students asked genAI include “what are the primary factors influencing elevator travel time,” “what is the maximum amount of g-force a human can feel vertically while still feeling comfortable,” and “can you help me understand this question without answering?” The results highlight the varied ways students utilize genAI tools in their problem-solving processes. While many students used it to directly generate solutions, others treated it more as a tutor and used it to support their problem-solving process rather than bypassing their own problem-solving.  This stark difference in the two types of use implies both the opportunities and challenges for the use of AI to support student learning.

\subsection{Perceived Helpfulness of GenAI Tools for STEM Problem-Solving}
After submitting their proposed prompts for genAI tools, students rated how helpful they think genAI tools would be in various aspects of STEM problem-solving. These aspects are informed by our previous research on distinct main aspects of STEM problem-solving and include: 1) providing explanations on the relevant physics knowledge underlying a problem; 2) collecting missing information; 3) assisting with calculations; and 4) verifying the correctness of a solution. Students rated genAI tools most helpful for explaining relevant physics knowledge, with 33 students (82.5\%) rating it as either helpful or highly helpful. Using genAI tools to assist with calculations was another aspect perceived positively, with 26 students (65\%) rating it helpful or highly helpful. In contrast, students expressed more uncertainty or skepticism about genAI’s utility in collecting missing information and validating the final solution. For collecting missing information, 24 students (60\%) were either unsure or rated the tools as unhelpful/very unhelpful. This skepticism was even more pronounced for verifying the correctness of a solution, with 28 students (70\%) expressing uncertainty or rating the tools as unhelpful/very unhelpful. 

\par
\begin{figure}[h]
\centering
\caption{Student responses to “How helpful do you think generative AI tools would be for the following tasks?” and instructor responses to “To what extent would you recommend/not recommend your students use generative AI tools like ChatGPT for the following problem-solving tasks, with the goal of enhancing their learning?”
}\label{fig4}
\includegraphics[width=\textwidth]{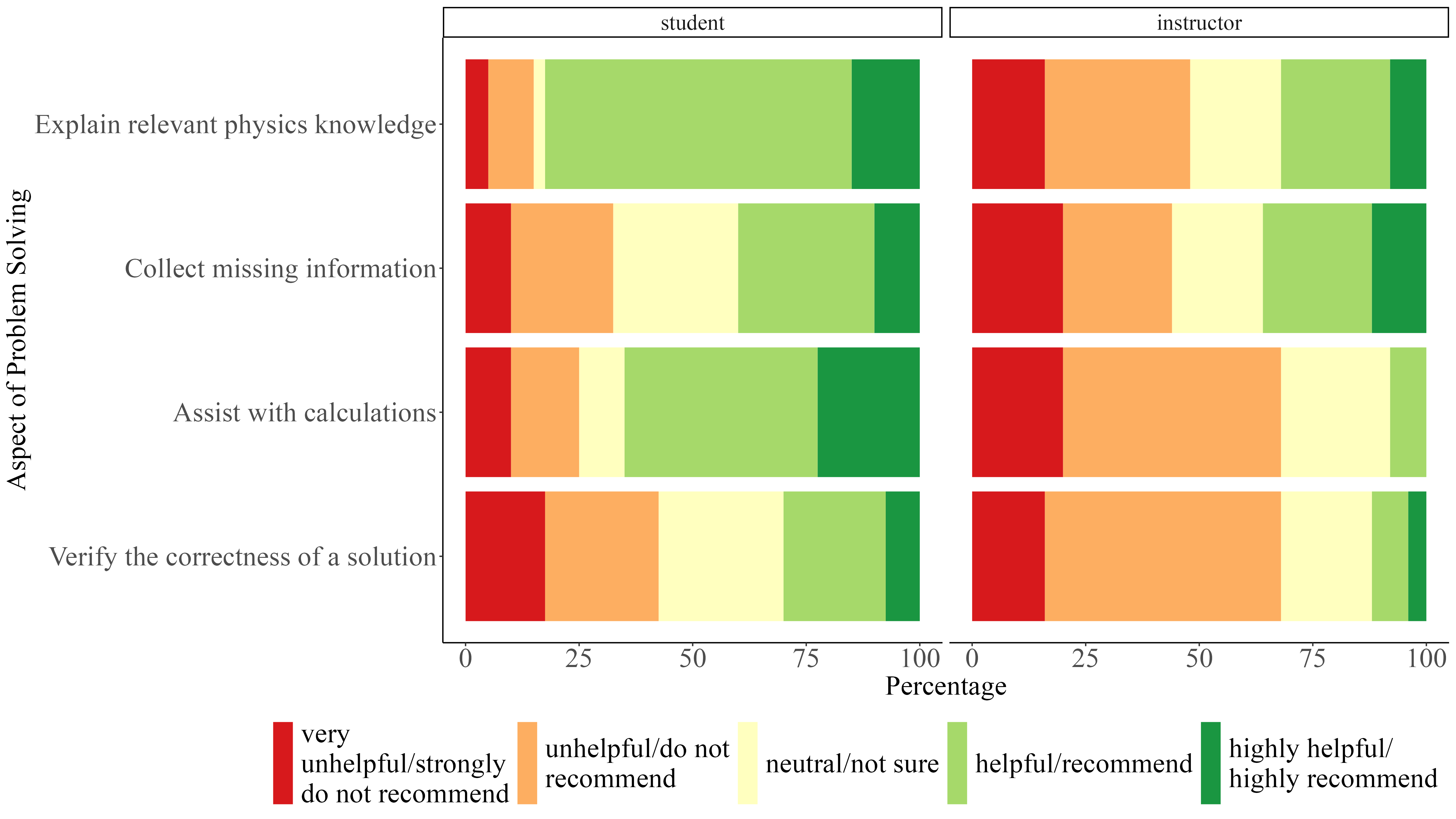}
\end{figure}

 While students generally viewed genAI tools favorably for explaining concepts and assisting with calculations, Instructors demonstrated more caution when it came to recommending how students should use genAI to support STEM problem-solving. They were shown the same physics problem and asked about recommending students using genAI tools for the same four aspects of problems-solving. Less than one-third of faculty respondents would recommend or highly recommend students using tools like ChatGPT for explaining concepts (n= 8, 29\%) or gathering missing information (n= 9, 32\%). Faculty showed even greater reservation for the other two aspects: only two (7\%) recommended using genAI tools for assisting with calculations, and three (11\%) recommended their use for verifying solution correctness. Fig.~\ref{fig4} summarizes the distribution of student and instructor ratings of genAI tools’ helpfulness in specific aspects of problem-solving. 

This stark contrast between faculty recommendations and students’ perceived helpfulness of genAI tools in explaining physics concepts and assisting with calculations highlights a potential misalignment in how these two groups view the role of genAI in supporting problem-solving in physics and potentially across STEM domains. On the other hand, faculty and students converged in their skepticism about genAI tools’ usefulness for checking the correctness of a solution, suggesting a shared recognition of the tools’ current limitations.

\subsection{Benefits and Risks of GenAI in STEM Education}
The final section of our survey included open-ended questions about the primary benefits and risks of using genAI tools in STEM education. We first examined the primary benefits by qualitatively analyzing instructor and student responses. Multiple codes were assigned to responses that mentioned more than one benefit. Both students and instructors identified a range of benefits of integrating genAI tools into STEM education. Notably, the instructor responses tended to cluster around a few key themes, while the student responses were more diverse. The most frequently cited benefit by both groups was that genAI tools can provide personalized learning support and resources, mentioned by 36\% of students and 50\% of faculty. For example, one student mentioned that “they (genAI tools) can assist with explanations to problems or help guide you in the right direction. They are good to help you understand the concepts in math and science courses that may be hard to follow for some.” Similarly, an instructor mentioned that “if used correctly, it could improve students’ conceptual understanding of topics, by making them ask ‘right’ questions in the prompt.” 

\par At the same time, students mentioned the benefits of using genAI to efficiently search and retrieve information (16\%) and summarize information (9\%), reflecting how they valued tools that can support their immediate learning needs and save time. For example, one student reported that “it also can get very descriptive in its answers that you simply cannot get with basic search engines.” Another reported that “it can give you a quick summary in an efficient manner in most cases, no matter the subject, you just have to feed it the information.” Additional benefits identified by students include support for specific tasks such as coding, writing, calculation, and brainstorming, as well as efficient handling of repetitive, mundane work. On the other hand, instructors in our survey highlighted genAI’s capacity to create novel learning and assessment materials (32\%) and support teaching preparation (32\%), reflecting their focus on instructional design and teaching. For example, several instructors proposed creating exercises that ask students to evaluate AI-generated solutions to physics problems, thereby challenging students to discern incorrect reasoning to enhance their learning. 

\par On the risk side, both students and instructors identified several key risks associated with using genAI tools in STEM education, with significant overlap in their major concerns. Misinformation emerged as the primary risk identified by both groups, cited by 67\% of faculty and 49\% of students. This concern is related to the potential inaccuracy of AI-generated content and the risk of creating misunderstandings among students. For instance, an instructor noted that “while ChatGPT has been demonstrated to sometimes be able to solve problems or provide explanations, it often provides total nonsense. For people freshly learning the concepts, I would be concerned about them only using this one source which is often completely incorrect and being unable to discern the difference.” Similarly, a student noted that “they are still inaccurate for a lot of stuff in my related field (biomed/microbiology). It doesn’t have the ability to fact check itself or check its calculations.” 

\par Both groups also highlighted the risk of genAI tools negatively impacting the quality of learning, although this was a much stronger concern of the instructors (50\% of instructors, 16\% of students). Instructors expressed particular concern about students’ over-reliance on these tools. As one faculty member put it, “students will become so dependent on these tools that they will think they no longer need to learn most material, as it can be accessed at any time using those tools. Students will fool themselves into thinking they have learned material which they have not.” Some students shared this concern. For example, one student response noted that “it is very easy to copy-paste an assignment description and have a chat bot produce reliable work. Aside from the ethical issues of academic dishonesty, this also prevents the student from learning the needed objectives to pass a course.” A third top risk noted by both groups is the risk to academic integrity (17\% of faculty, 25\% of students). Interestingly, students’ responses in this category revealed a dual concern, that they recognized both the risk of using genAI to cheat and the risk of being accused of academic dishonesty by their instructors and schools when using genAI to help them with assignments or edit their writings in what they thought was an ethical way. 

\section{Discussion}
The findings of this study provide important insights into the current landscape of genAI tool use by college students in their STEM coursework. First, our survey revealed a high adoption rate of genAI tools among students, with 37 out 40 participants reporting using predominantly free versions of LLM tools. Second, students utilized these tools for a range of tasks in their STEM coursework, including finding explanations, exploring related topics, summarizing readings, and helping with problem-set questions. 

\par The high adoption rate of genAI tools among students indicate students’ willingness to engage with this emerging technology. At the same time, students’ reliance on free versions of these tools has significant implications for STEM education. Specifically, students’ prevalent use of genAI for finding explanations for STEM topics and solutions to STEM problems poses a risk of being exposed to and reinforcing misunderstandings if the AI-generated content is inaccurate. This risk is potentially exacerbated due to students’ use of free versions of LLMs, which may not always incorporate the latest improvement in model performance regarding factual accuracy and up-to-date information. Moreover, the accessibility gap between free and paid versions of genAI tools raises educational equity concerns, as students with access to paid versions may receive better information, potentially gaining an academic advantage over those limited to free versions. Consequently, colleges should consider providing free access to reliable, education-focused genAI tools for all students. Additionally, there is a pressing need to develop students’ critical evaluation mindsets and skills when working with AI-generated information, for example, by cross referencing AI-generated explanations with textbooks and instructor-provided materials. 

\par Another key finding from this study is the divergent approaches students took when utilizing genAI to support STEM problem-solving. We found a near-even split between students who prompted a genAI-based chatbot to directly solve the sample problem and those who asked specific questions to scaffold their own problem-solving process. This result highlights both the varying levels of prompt engineering skills among students and the differences in their epistemic beliefs about genAI tools’ role in supporting their STEM problem-solving and learning. Students' tendency to use generative AI as a shortcut for direct solutions, rather than as a scaffold for independent problem-solving, mirrors their preference for passive lectures over active learning experiences \citep{deslauriers2019measuring}. On one hand, students who rely on these tools to directly generate solutions for problems in their coursework may feel that they have learned how to solve the problem after studying AI-generated solutions with little mental effort. However, this approach bypasses deeper learning and hinders the development of STEM problem-solving competency, similar to how passive lectures can create a false sense of fluency yet lead to less actual learning. On the other hand, students who use genAI as a scaffold must expend more mental efforts to solve problems on their own, but this process can help deepen their conceptual understanding and develop effective problem-solving skills, mirroring the benefits of active learning. 

\par Future research should investigate how students’ beliefs about genAI’s role in problem-solving and learning influence the ways they use genAI tools and their subsequent learning outcomes.  Studies could explore whether students who use AI to provide scaffolds rather than to generate solutions achieve better learning outcomes and develop more effective problem-solving skills. Additionally, research needs to examine how institutions and course instructors can educate students about productive AI use strategies and help them develop effective prompt engineering skills with the goal of enhancing learning.  

\par One promising finding from the study is that students demonstrated reasonable assessments of genAI tools’ helpfulness for different aspects of STEM problem-solving. Students rated these tools most helpful for explaining relevant domain knowledge underlying a problem and least helpful for verifying solution correctness. Students’ high rating for explaining domain knowledge aligns with our previous research evaluating LLMs’ capability to independently solve complex STEM problems \citep{wang2024examining}. This indicates that students are, to some extent, aware of the tools’ affordances and limitations in problem-solving contexts. At the same time, the survey revealed a divergence between students’ perceived helpfulness and instructor recommendations regarding genAI tools usage in problem-solving. Less than one-third of the instructors recommended using genAI tools to support any aspect of problem-solving. This misalignment calls for ongoing dialogue between students and instructors about the appropriate and effective use of genAI tools in STEM courses, discussing when they support learning versus potentially enabling cheating. We also recommend that STEM educators empirically evaluate the potential and limitations of genAI in solving problems specific to their coursework to provide more grounded and targeted recommendations to students. 

\par Lastly, our qualitative analyses of open-ended responses regarding the benefits of using genAI tools in college STEM courses revealed both shared and distinct perspectives among students and instructors. The primary benefit identified by both groups was the provision of personalized learning support and resources. Instructors also highlighted instructional benefits such as genAI’s potential for creating novel learning and assessment materials and aiding teaching preparation. On the other hand, students emphasized efficiently searching, retrieving, and summarizing information as another key benefit. The focus on efficiency is a recurring theme in students’ use and perception of genAI tools, as evidenced by their top reported reason for using these tools (to save time) and in their prompting strategy of directly generating solutions when using genAI-based chatbot to solve problems. This trend aligns with broader patterns in students’ use of digital technologies in higher education. \cite{henderson2015students} found that college students primarily valued digital technologies for their ability to support organizational tasks, time management, and efficient completion of coursework. The authors concluded that while digital technologies have become central to student experiences, they are not transforming the nature of university teaching and learning in the ways often envisioned by educators and researchers. GenAI tools, despite their potential, may be following a similar trajectory for the time being—their use for quick solutions and time-saving purposes may not necessarily lead to enhanced learning outcomes or the development of higher-order competencies like problem-solving. 

\par In conclusion, this study finds that college students are already using genAI tools in their STEM coursework. At the same time, our results highlight two critical needs. First, there is a need to measure and teach students AI literacy for more effective use of these tools to enhance their learning. Second, there is a need for more thoughtful design of genAI-based tools to enhance learning outcomes. One direction for future technology development is to design genAI-based tutors that can guide students through problem-solving processes while avoiding providing direct solutions. For instance, promising work by \cite{kestin2024ai} in an introductory college physics class demonstrated the potential of carefully designed genAI-based tutors in enhancing student engagement and learning outcomes while saving time. Such tutors can be designed to leverage personalized scaffolding and feedback to ensure that students’ cognitive efforts remain within a reasonable range, preventing excessive frustration and making active learning experiences more effective for students with different levels of background knowledge. As genAI continues to develop and evolve, it is crucial to ensure that the technology supports genuine learning rather than inadvertently creating dependencies that may hinder students’ development of critical knowledge and competencies. 

\section{Limitations}
There are several limitations associated with this study that warrant consideration. First, our relatively small sample size of 40 college students and 28 instructors may limit the generalizability of our findings. However, it’s important to note that participants came from diverse institutions across the US, which mitigates the potential bias that could arise from the academic policies or culture of any single institution. This institutional diversity enhances the breadth of perspectives captured in our study, even with limited sample size. Nonetheless, the patterns and insights identified, particularly regarding specific uses of genAI tools, would benefit from further validation through large-scale investigations. Second, our study relied on voluntary participation through an online platform, which may have introduced selection bias. Students who chose to participate in our survey might have been more familiar with or interested in genAI tools, potentially skewing our results towards higher adoption rates and more positive attitudes. As a result, this sample may not be representative of the entire US college student population studying STEM subjects. Lastly, the study focused on STEM courses with a particular emphasis on problem-solving. This focus limits the applicability of our findings to non-STEM disciplines, each of which may have unique challenges and opportunities for genAI integration. Despite these limitations, this study provides valuable insights into the use of genAI tools among college students by highlighting current use patterns, perceptions, and implications for STEM education.

\bibliography{sn-bibliography}

\end{document}